\documentclass[12pt,reqno]{amsart}
\allowdisplaybreaks[4]
\usepackage{graphicx} 
\usepackage{amssymb}
\usepackage{amsmath}
\usepackage{cite}

\numberwithin{equation}{section}
\makeatletter      
\@addtoreset{equation}{section}
\makeatother       
\newtheorem{proposition}{Proposition}[section]
\newtheorem{corollary}{Corollary}[section]
\newtheorem{lemma}{Lemma}[section]

\begin{document}

\title[Additional Symmetries and String Equation]{Additional Symmetries and String Equation of
the CKP hierarchy}

\author{Jingsong He\dag\ddag, Kelei Tian\dag, Angela Foerster\ddag and  Wen-xiu Ma\S}
\dedicatory { \dag\ Department of Mathematics, USTC, Hefei, 230026 Anhui, P.\ R.\ China\\
\ddag\ Instituto de F\'{\i}sica da UFRGS, Av. Bento Gon\c{c}alves 9500, Porto
Alegre, RS - Brazil\\
\S Department of Mathematics, University of South Florida
Tampa, FL 33620-5700, USA
}

\begin{abstract}
Based on the Orlov and Shulman's M operator,  the additional
symmetries and the string equation of the CKP hierarchy are
established, and then the higher order constraints on $L^l$ are
obtained. In addition, the generating function and  some properties
are also given. In particular, the additional symmetry flows form a
new infinite dimensional algebra $W^C_{1+\infty}$, which is a
subalgebra of $W_{1+\infty}$.
\end{abstract}

 \maketitle

\keywords{Keywords:\ CKP hierarchy,\ additional symmetries,\ string
equation}

 Mathematics Subject Classification(2000):\ 17B80,\ 37K05,\ 37K10

PACS(2003):\ 02.30.Ik


\section{Introduction}
Since its introduction in 1980, the  Kadomtsev-Petviashvili(KP) hierarchy \cite{dkjm,dl1} is
one of the most important research topics  in the area of classical integrable systems. In particular, the
study of its  symmetries plays a central role in the development of  this  theory.
In this context, additional symmetries, which correspond to  a special kind of symmetries depending explicitly on the independent variables $t_n$ of the KP hierarchy,  have been analyzed
using two different approaches. In the first one, the explicit form of the additional
symmetry flows (action on the wave function, or equivalently on the Lax operator $L$) of the KP hierarchy
was given by Orlov and Shulman(OS) \cite{os1} through a novel operator $M$,
which can be used to form a centerless $W_{1+\infty}$ algebra. Actually, these results go
back to some previous works on  the $t$ (time variable) and $x$ (space variable) dependent
symmetries of the KP equation, which forms an infinite dimensional Lie algebra, founded by
Oevel and Fuchssteiner\cite{of1,f1}, and Chen, Lee and Lin\cite{cll1}. In the second approach,
there exists the Sato B$\ddot{a}$cklund symmetry defined by the vertex operator $X(\lambda,\mu)$  acting
on the $\tau$ function of the KP hierarchy, with $X(\lambda,\mu)$  serving as a generating
function of the $W_{1+\infty}$ algebra\cite{dkjm}. It is quite surprising that no direct
connection was realized between these two types of symmetries for a long period of time. But,
in fact, the $W_{1+\infty}$-algebra of  additional symmetry flow defined by OS can be lifted to
its central extension by acting on the $\tau$ function of the KP hierarchy\cite{asv1,asv2,asv3}.
In this process, the Adler-Shiota-van Moerbeke(ASvM) formula plays a crucial role. Almost at the same time, Dickey
presented a very elegant and compact proof of the ASvM formula\cite{dl3}. He also found the
action of the additional symmetries on the Grassmannian and gave a straightforward derivation
of the action of the additional symmetries  on  the $\tau$-functions\cite{dl4}.

It is well known that there are two kinds of sub-hierarchies of KP,
a BKP hierarchy\cite{dkjm} and a CKP hierarchy \cite{dkjm2}. For the
BKP hierarchy, its Virasoro constraints and the ASvM formula have
been constructed by Johan van de Leur\cite{vdl1,vdl2} using  an algebraic
formalism. Very recently, an alternative proof of the ASvM formula of the
BKP hierarchy was given by Tu\cite{tu1} by means of Dickey's method
\cite{dl3}. So, it would be natural to ask if some corresponding
results related to the additional symmetries in the CKP hierarchy
also applies. The question concerning the similarities and
differences between the BKP and CKP hierarchies is also very
relevant in this scenario. Obviously, in contrast to the BKP
hierarchy, we can not find an ASvM formula for CKP because this
hierarchy does not possess a sole $\tau$-function. However this
fact does not destroy the existence of additional
symmetries and the string equation for the CKP hierarchy, as we
shall deliver. The main purpose of this article is to construct the
additional symmetries and string equations for the CKP hierarchy
providing an answer to all these relevant questions.

The organization of this paper is as follows. We recall some basic facts for the CKP hierarchy in
section 2. The OS's $M$ operator, additional symmetry and string equation are discussed in
sections 3,4 and 5, respectively. Section 6 is devoted to conclusions and discussions.

\section{CKP Hierarchy}
Let $L$ be the pseudo-differential operator,
\begin{equation}\label{KPlaxoperator}
L=\partial +u_1\partial^{-1}+u_2\partial^{-2}+
u_3\partial^{-3}+\cdots,
\end{equation}
and then the KP hierarchy is  defined by the set of partial differential equations $u_i$
with respect to independent variables $t_j$
\begin{equation}\label{KPhierarchy}
\dfrac{\partial L}{\partial {t_n}}=[B_n, L],\ n=1,2,3, \cdots.
\end{equation}
Here $B_n=(L^n)_+=\sum\limits_{k=0}^n a_k\partial^k$ denotes the
non-negative powers of $\partial $ in $L^n$, $\partial
=\partial/\partial_x$, $u_i=u_i(x=t_1,t_2,t_3,\cdots,)$. The other
notation $L^n_{-}=L^n-L^n_+$ will be needed by the sequent text. $L$
is called the Lax operator and eq.(\ref{KPhierarchy}) is called the
Lax equation of the KP hierarchy. In order to define the CKP
hierarchy, we need a formal adjoint operation $*$ for an arbitrary
pseudo-differential operator $P=\sum\limits_i p_i \partial^i$,
$P^*=\sum\limits_i (-1)^i\partial^i  p_i$. For example,
$\partial^*=-\partial$, $(\partial^{-1})^*=-\partial^{-1}$, and
$(AB)^*=B^*A^*$ for two operators. The CKP hierarchy\cite{dkjm2} is
a  reduction of  the KP hierarchy by the constraint
\begin{equation}\label{CKPconstraint}
L^*=- L ,
\end{equation}
which compresses all even flows of the KP hierarchy, i.e. the Lax equation of the CKP hierarchy has only odd flows ,
\begin{equation}\label{CKPhierarchy}
\dfrac{\partial L}{\partial {t_{2n+1}}}=[B_{2n+1}, L],\ n=0,1,2, \cdots.
\end{equation}
Thus $u_i=u_i(t_1,t_3, t_5,\cdots)$ for the CKP hierarchy.

  The Lax equation of the CKP hierarchy can be given by the consistent conditions of the following
set of linear partial differential equations
\begin{equation}\label{CKPlinearsystem}
Lw(t,\lambda)=\lambda w(t,\lambda),\dfrac{\partial w(t,\lambda)}{\partial {t_{2n+1}}}
=B_{2n+1}w(t,\lambda), \ t=(t_1,t_3,t_5, \cdots).
\end{equation}
Here $w(t,\lambda)$ is identified as a wave function. Let $\phi$
be the wave operator(or Sato operator) of the CKP hierarchy
$\phi=1+\sum_{i=1}^{\infty}w_i\partial^{-i}$, then
the Lax operator and the wave function admit the following
representation
\begin{equation}\label{satorepresenation}
L=\phi \partial \phi^{-1}, \ \ w(t,\lambda)=\phi(t)e^{\xi(t,\lambda)}=\hat{w}e^{\xi(t,\lambda)},
\end{equation}
in which $\xi(t,\lambda)=\lambda t_1 +\lambda^3t_3+\cdots+\lambda^{2n+1}t_{2n+1}+\cdots$,
$\hat{w}=1+\frac{w_1}{\lambda}+\frac{w_2}{\lambda^2}+\frac{w_3}{\lambda^3}+\cdots$.
It is easy to show that the Lax equation is equivalent to Sato equation
\begin{equation}\label{satoequation}
\dfrac{\partial \phi}{\partial{t_{2n+1}}}= -L^{2n+1}_{-} \phi,
\end{equation}
and the constraint on $L$ in eq.(\ref{CKPconstraint}) is transformed to
the constraint on the wave operator
\begin{equation}\label{CKPconstraintwaveoperator}
\phi^*=\phi^{-1}.
\end{equation}
The eq.(\ref{CKPconstraintwaveoperator}) is a crucial condition to construct the additional
symmetries of the CKP hierarchy, which will affect the action of the additional symmetry on $\phi$.
It leads to a distinct definition of the additional symmetry
in comparison to the cases of the KP and BKP hierarchies.

\section{Orlov-Schulman's M operator and its adjoint}
The Orlov and Shulman's $M$ operator\cite{os1} is also applicable to construct the additional
symmetries of the CKP hierarchy when  all of the even independent variables $t_{2n}$ are frozen. Thus define
\begin{equation}\label{Moperator}
M=\phi\Gamma \phi^{-1},\  \Gamma=\sum\limits_{i=0}^{\infty}(2i+1)t_{2i+1}\partial^{2i}=t_1+3t_3\partial^2
+5t_5\partial^4+\cdots.
\end{equation}
A direct calculation shows that the operator $M$ satisfies
\begin{equation}\label{flowsofM}
[L,M]=1, \; \partial_{t_{2n+1}}M=[B_{2n+1},M], \; Mw(t,z)=\left(\partial_zw(t,z)\right).
\end{equation}
Further,
\begin{equation}\label{flowsofMm}
\dfrac{\partial M^m}{\partial t_{2n+1}}=[B_{2n+1},M^m],\;\dfrac{\partial M^mL^l}{\partial t_{2n+1}}=
[B_{2n+1},M^mL^l]
\end{equation}
with the help of $\frac{\partial L^l}{\partial t_{2n+1}}=[B_{2n+1},L^l]$.
Moreover, on the space of wave functions $w(t,z)$, $[L,M]=1$ and $[z,\partial_z]=-1$ induce an
anti-isomorphic between $(L,M)$ and $(z,\partial_z)$, and
\begin{equation}\label{generalLM}
M^mL^lw(t,z)=z^l\left(\partial_z^mw(t,z)\right),\; L^lM^mw(t,z)=\partial_z^m\left(z^l w(t,z)\right),
m,l\in Z,m\ge 0.
\end{equation}

On the other hand, we need some information of the adjoint wave function $w^*$(which is the wave function of
the adjoint system $L^*$)  and  $M^*$(the formal adjoint of $M$). For the CKP hierarchy,
we have
\begin{equation}\label{adjointw}
w^*(t,z)=(\phi^*)^{-1}e^{-\xi(t,z)}=\phi e^{\xi(t,-z)}=w(t,-z),
\end{equation}
and
\begin{equation}\label{adjointM}
M^*=\left(\phi \Gamma\phi^{-1} \right)^*=(\phi^{-1})^* \Gamma^* \phi^*=M
\end{equation}
by using eq.(\ref{CKPconstraintwaveoperator}), $\Gamma^*=\Gamma$ and $\xi(t,-z)=-\xi(t,z)$.
Furthermore, $L^*$ and $M^*$ satisfy
\begin{equation}\label{adjointMandL}
[L^*,M^*]=[-L,M]=-1,
\end{equation}
and
\begin{equation}\label{onadjointfunction}
L^*w^*(t,z)=zw^*(t,z),\;\partial_{t_{2n+1}}w^*(t,z)=-B_{2n+1}^*w^*(t,z),
\;M^*w^*(t,z)=-\partial_zw^*(t,z).
\end{equation}

\section{Additional Symmetries}
We are now in a position to define the additional flows, and then to prove that they are
symmetries, which are called additional symmetries of the CKP hierarchy. Similar to the case of the BKP\cite{tu1},
we introduce additional independent variables $t^*_{m,l}$ and define the action of the additional flows
on the wave operator as
\begin{equation}\label{definitionadditionalflowsonphi}
\dfrac{\partial \phi}{\partial {t^*_{m,l}}}=-\left(A_{m,l}\right)_{-}\phi,
\end{equation}
where $A_{m,l}=A_{m,l}(L,M)$ are monomials in $L$ and $M$ and their explicit forms are undetermined.
\begin{proposition}
The additional flows act on $L$ and $M$ as
\begin{equation}\label{CKPadditionalflowsonLandM}
\dfrac{\partial L}{\partial{t^*_{m,l}}}=-[(A_{m,l})_{-}, L],\;
 \dfrac{\partial M}{\partial{t^*_{m,l}}}=-[(A_{m,l})_{-}, M]
\end{equation}
\end{proposition}
\noindent \textbf{Proof} By performing the derivative on $L$(\ref{satorepresenation}) and using
eq.(\ref{definitionadditionalflowsonphi}), we get
\begin{eqnarray*}
(\partial_{t^*_{m,l}}L)&=&(\partial_{t^*_{m,l}}\phi)\ \partial \phi^{-1}
+ \phi\ \partial\ (\partial_{t^*_{m,l}}\phi^{-1})\\
&=&-(A_{m,l})_{-} \phi\ \partial\ \phi^{-1}- \phi\ \partial \phi^{-1}\ (\partial_{t^*_{m,l}}\phi)
\ \phi^{-1}\\
&=&-(A_{m,l})_{-} L+ L (A_{m,l})_{-}=-[(A_{m,l})_{-}, L].
\end{eqnarray*}
For the action on $M$ given in eq.(\ref{Moperator}), there exists similar derivation as $(\partial_{t^*_{m,l}}L)$,i.e.
\begin{eqnarray*}
(\partial_{t^*_{m,l}}M)&=&(\partial_{t^*_{m,l}}\phi)\ \Gamma \phi^{-1}
+ \phi\ \Gamma\ (\partial_{t^*_{m,l}}\phi^{-1})\\
&=&-(A_{m,l})_{-} \phi\ \Gamma\ \phi^{-1}- \phi\ \Gamma \phi^{-1}\ (\partial_{t^*_{m,l}}\phi)
\ \phi^{-1}\\
&=&-(A_{m,l})_{-} M+ M (A_{m,l})_{-}=-[(A_{m,l})_{-}, M].
\end{eqnarray*}
Here the fact that $\Gamma$ does not depend on the additional flows variables $t^*_{m,l}$ has been used.
\hfill $\square$
\begin{corollary}\label{additionflowsonLnMmAnk}
\begin{equation}\label{eqadditionflowsonLnMmAnk}
\dfrac{\partial L^n}{\partial{t^*_{m,l}}}=-[(A_{m,l})_{-}, L^n],\;
 \dfrac{\partial M^m}{\partial{t^*_{m,l}}}=-[(A_{m,l})_{-}, M^m],\;
 \dfrac{\partial A_{n,k}}{\partial{t^*_{m,l}}}=-[(A_{m,l})_{-}, A_{n,k}]
\end{equation}
\begin{equation}
\dfrac{\partial A_{n,k}}{\partial{t_{2n+1}}}=[B_{2n+1}, A_{n,k}]
\end{equation}
\end{corollary}
\noindent \textbf{Proof} We present here only the proof of the first equation. The others
can be proved in a similar way. The derivative of $L^n$ with respect to $t^*_{m,l}$ leads to
\begin{eqnarray*}
\dfrac{\partial L^n}{\partial{t^*_{m,l}}}=\dfrac{\partial L}{\partial{t^*_{m,l}}}L^{n-1}+
L \dfrac{\partial L}{\partial{t^*_{m,l}}} L^{n-2}+\cdots
+L^{n-2} \dfrac{\partial L}{\partial{t^*_{m,l}}} L
+L^{n-1} \dfrac{\partial L}{\partial{t^*_{m,l}}}=\sum\limits_{k=1}^n L^{k-1}
\dfrac{\partial L}{\partial{t^*_{m,l}}} L^{n-k}
\end{eqnarray*}
and then taking $\dfrac{\partial L}{\partial{t^*_{m,l}}}=-[(A_{m,l})_{-}, L]$ into the above formula,
which is followed by
\begin{eqnarray*}
\dfrac{\partial L^n}{\partial{t^*_{m,l}}}=-\sum\limits_{k=1}^n L^{k-1}
[(A_{m,l})_{-}, L]
 L^{n-k}=-[(A_{m,l})_{-}, L^n]. \text{\hspace{5cm}} \square
\end{eqnarray*}
\begin{proposition}
The additional flows $\dfrac{\partial }{\partial{t^*_{m,l}}}$ commute with the CKP hierarchy flows
$\dfrac{\partial }{\partial{t_{2n+1}}}$, i.e.
\begin{equation}
[\partial_{t^*_{m,l}}, \partial_{t_{2n+1}}]=0
\end{equation}
Thus they are symmetries of the CKP hierarchy. Here
$\partial_{t^{*}_{m,l}}=\frac{\partial}{\partial{t^{*}_{m,l}}},
\partial_{t_{2n+1}}=\frac{\partial}{\partial{t_{2n+1}}}$.
\end{proposition}
\noindent \textbf{Proof} According to the definition,
\begin{equation*}
[\partial_{t^*_{m,l}},\partial_{t_{2n+1}}]\phi=\partial_{t^*_{m,l}} (\partial_{t_{2n+1}}\phi)-
\partial_{t_{2n+1}} (\partial_{t^*_{m,l}}\phi),
\end{equation*}
and using the action of the additional flows and the CKP flows on $\phi$, then
\begin{eqnarray*}
[\partial_{t^*_{m,l}},\partial_{t_{2n+1}}]\phi&=&
-\partial_{t^*_{m,l}}\left(L^{2n+1}_{-}\phi\right) +
\partial_{t_{2n+1}} \left((A_{m,l})_{-}\phi \right)\\
&=&-(\partial_{t^*_{m,l}}L^{2n+1} )_{-}\phi-
(L^{2n+1})_{-}(\partial_{t^*_{m,l}}\phi)+ (\partial_{t_{2n+1}} A_{m,l})_{-}\phi
+ (A_{m,l})_{-}(\partial_{t_{2n+1}}\phi).
\end{eqnarray*}
By the corollary \ref{additionflowsonLnMmAnk} and eq.(\ref{satoequation}), it equals
\begin{eqnarray*}
[\partial_{t^*_{m,l}},\partial_{t_{2n+1}}]\phi&=&[(A_{m,l})_{-}, L^{2n+1}]_{-}\phi+
(L^{2n+1})_{-}(A_{m,l})_{-}\phi\\
&+&[(L^{2n+1})_{+},A_{m,l}]_{-}\phi-(A_{m,l})_{-}(L^{2n+1})_{-}\phi\\
&=&[(A_{m,l})_{-}, L^{2n+1}]_{-}\phi- [(A_{m,l})_{-}, L^{2n+1}_{+}]_{-}\phi+
[L^{2n+1}_{-},(A_{m,l})_{-}]\phi\\
&=&[(A_{m,l})_{-}, L^{2n+1}_{-}]_{-}\phi+ [L^{2n+1}_{-},(A_{m,l})_{-}]\phi=0
\end{eqnarray*}
In the second equality of the above derivation, $[L^{2n+1}_{+}, (A_{m,l})]_{-}=
[L^{2n+1}_{+}, (A_{m,l})_{-}]_{-}$ has been used, since $[L^{2n+1}_{+}, (A_{m,l})_{+}]_{-}=0$.
The last equality holds because $(P_{-})_{-}=P_{-}$ for arbitrary pseduo-differential operator
$P$.
\hfill $\square$

In contrast to the KP hierarchy, it is not possible to let $A_{m,l}=M^mL^l$, because there is a
crucial restriction on $\phi$ (or on L equivalently ) in eq.(\ref{CKPconstraintwaveoperator}).
This restriction results in the following sufficient condition for $A_{m,l}$.
\begin{proposition}\label{additionalsymmetrygenerator}
For the CKP hierarchy, it is  sufficient to ask
\begin{equation}\label{constraintonAml}
A_{m,l}^*=-A_{m,l}.
\end{equation}
Thus we can let
\begin{equation}\label{additionalsymmeterygenerator}
A_{m,l}=M^mL^l-(-1)^lL^lM^m
\end{equation}
\end{proposition}
\noindent\textbf{Proof} The action of the additional flows $\partial_{t^*_{m,l}}$
on the adjoint wave
operator $\phi^*$ can be obtained by two different ways. The first is to do a formal adjoint operation on
eq. (\ref{definitionadditionalflowsonphi}),
\begin{equation}\label{additionalflowsonadjointphia}
\partial_{t^*_{m,l}}\phi^*=-\phi^*( A_{m,l})_{-}^*.
\end{equation}
The second is to do a derivative with respect to $t^*_{m,l}$ on $\phi^*$ and use the constraint
relation $\phi^{*}=\phi^{-1}$,
\begin{equation}\label{additionalflowsonadjointphib}
\partial_{t^*_{m,l}}\phi^*=\partial_{t^*_{m,l}}\phi^{-1}=-
\phi^{-1}(\partial_{t^*_{m,l}}\phi)\phi^{-1}=\phi^{-1}(A_{m,l})_{-}=\phi^{*}(A_{m,l})_{-}.
\end{equation}
By comparing eq.(\ref{additionalflowsonadjointphia}) and eq.(\ref{additionalflowsonadjointphib}),
we have
\begin{equation}\label{proofcosstraintonAm1}
(A_{m,l})_{-}=-(A^*_{m,l})_{-},
\end{equation}
and thus it is sufficient to let $A_{m,l}=-A^*_{m,l}$.  Moreover, $(M^mL^l)^*=(L^l)^*(M^m)^*
=(-1)^lL^lM^m$, so $A_{m,l}= M^mL^l-(-1)^lL^lM^m$ satisfies the requirement of
eq.(\ref{proofcosstraintonAm1}).
\hfill$\square$\\
\noindent \textbf{Remark:} Here the generators of the additional
symmetries  $ A_{m,l}$ are also different with the counterparts of
the  BKP hierarchy, which has different constraints on
$\phi$\cite{kt1,tu1}.
\begin{proposition}\label{WinfiniteCalgebra}
Acting on the space of the wave operator $\phi$, $\partial_{t^*_{m,l}}$ forms new centerless
$W^C_{1+\infty}$- subalgebra of centerless $W_{1+\infty}$.
\end{proposition}
\noindent
\textbf{Proof}
By using  $A_{n,k}^*=-A_{n,k}$, it is easy to compute
\begin{eqnarray}
([A_{m,l}, A_{n,k}])^*=A_{n,k}^*A_{m,l}^*-A^*_{m,l}A_{n,k}^*\notag\\
=A_{n,k}A_{m,l}-A_{m,l}A_{n,k}=-[A_{m,l}, A_{n,k}],
\end{eqnarray}
which shows that they form a close set under the commutator operation.
Therefore it can be expanded by other generator,
\begin{equation}
[A_{m,l}, A_{n,k}]=\sum\limits_{p,q}C^{pq}_{nk,ml}A_{p,q},
\end{equation}
and so
\begin{equation}\label{negativepartcommutatorAmlAnk}
[A_{m,l}, A_{n,k}]_{-}=\sum\limits_{p,q}C^{pq}_{nk,ml}(A_{p,q})_{-}.
\end{equation}
This fact will be used in the computation of commutator of $\partial_{t^*_{m,l}}$. Now we start
to do this.
By using  eq.(\ref{definitionadditionalflowsonphi}),
\begin{eqnarray*}
&[\partial_{t^*_{m,l}},\partial_{t^*_{n,k}}]\phi=
\partial_{t^*_{m,l}}(\partial_{t^*_{n,k}}\phi)-
\partial_{t^*_{n,k}}(\partial_{t^*_{m,l}}\phi)
=-\partial_{t^*_{m,l}}\left((A_{n,k})_{-}\phi\right)
+\partial_{t^*_{n,k}}\left((A_{m,l})_{-}\phi\right)\\
&=-(\partial_{t^*_{m,l}} A_{n,k})_{-}\phi-(A_{n,k})_{-}(\partial_{t^*_{m,l}} \phi)+
(\partial_{t^*_{n,k}} A_{m,l})_{-}\phi+ (A_{m,l})_{-}(\partial_{t^*_{n,k}} \phi).
\end{eqnarray*}
On account of eq.(\ref{definitionadditionalflowsonphi}) again and
 eq.(\ref{eqadditionflowsonLnMmAnk}),
 \begin{eqnarray*}
[\partial_{t^*_{m,l}},\partial_{t^*_{n,k}}]\phi &=&
[(A_{m,l})_{-},A_{n,k}]_{-}\phi+(A_{n,k})_{-}(A_{m,l})_{-}\phi-
[(A_{n,k})_{-},A_{m,l}]_{-}\phi-(A_{m,l})_{-}(A_{n,k})_{-}\phi\\
&=&[A_{m,l},A_{n,k}]_{-}\phi.
 \end{eqnarray*}
Taking eq.(\ref{negativepartcommutatorAmlAnk}) into the above formula, and then using
eq.(\ref{definitionadditionalflowsonphi}) inversely,
 we get
\begin{eqnarray*}
&[\partial_{t^*_{m,l}},\partial_{t^*_{n,k}}]\phi=-
\sum\limits_{p,q}
C^{pq}_{nk,ml}(A_{p,q})_{-}\phi=\sum\limits_{p,q}
C^{pq}_{nk,ml}(\partial_{t^{*}_{p,q}}\phi ),
\end{eqnarray*}
which is equivalent to
\begin{equation*}
\mbox{\hspace{3cm}}
[\partial_{t^*_{m,l}},\partial_{t^*_{n,k}}]=\sum\limits_{p,q}
C^{pq}_{nk,ml}\partial_{t^{*}_{p,q}}. \mbox{\hspace{7cm}}\square
\end{equation*}

To have a better understanding of the additional symmetry flows, we provide several
typical examples.
\begin{corollary}
From eq (\ref{additionalsymmeterygenerator}) we get $A_{m,1}=2LM^m-mM^{m-1},m\ge 1,m\in Z$, thus
 the corresponding flows on L are
\begin{equation}\label{tm1onL}
\dfrac{\partial L}{\partial {t^*_{m,1}}}=[-2(LM^m)_{-}+m(M^{m-1})_{-},L].
\end{equation}
Let $m=1$,
\begin{equation}\label{t11onL}
\dfrac{\partial L}{\partial{t^*_{1,1}}}=-[2(LM)_{-},
L]=2L+2[(LM)_{+}, L]
\end{equation}
infers
\begin{equation}\label{t11onui}
\dfrac{\partial u_i}{\partial{t^*_{1,1}}}=
2\left(x\dfrac{\partial u_i}{\partial x}+(i+1)u_i + \sum\limits_{j=1}^{\infty}
(2j+1)t_{2j+1}\dfrac{\partial u_i}{\partial_{t_{2j+1}}}
\right), \; i=1,2,3,\cdots.
\end{equation}
\end{corollary}
\noindent\textbf{Proof}
Eqs.(\ref{tm1onL}) and(\ref{t11onL}) are obtained by definition.
For the eq.(\ref{t11onui}), first of all, we should find $LM$, which is expressed by
\begin{eqnarray}\label{LM}
LM=\phi \partial \Gamma \phi^{-1}=\phi\partial x \phi^{-1}+
 \sum\limits_{i=1}^{\infty}(2i+1)t_{2i+1}\phi \partial^{2i+1}
\phi^{-1}
\end{eqnarray}
with the help of  eq.(\ref{satorepresenation}) and eq.(\ref{Moperator}).
Furthermore,
by $\partial x=x\partial +1$ and $\partial^{-i} x=x\partial^{-i}-i\partial^{-i-1}$,
$$
\phi \partial x \phi^{-1}=(x\partial +1+w_1 x)\phi^{-1}+\left(
\sum\limits_{i=2}^{\infty}\left(w_i
x\partial^{-(i-1)}-iw_i\partial^{-i} \right)\right)\phi^{-1},
$$
and then
\begin{equation}\label{phipartialxphiinverse}
(\phi \partial x \phi^{-1})_+=x\partial +1.
\end{equation}
Here the $\phi^{-1}=1-w_1\partial^{-1}+\cdots$ is used.
Taking eq.(\ref{phipartialxphiinverse}) into $LM$, we have
\begin{equation}\label{LM+}
(LM)_{+}=x\partial+1+ \sum\limits_{i=1}^{\infty}(2i+1)t_{2i+1}L^{2i+1}_{+}.
\end{equation}
Taking eq.(\ref{LM+}) back into eq.(\ref{t11onL}),
\begin{eqnarray}
\partial_{t^*_{1,1}}L=2L+2[x\partial,L]+2[\sum\limits_{i=1}^{\infty}(2i+1)t_{2i+1}L^{2i+1}_+, L]\notag\\
=2L+ 2[x\partial,L]+2\sum\limits_{i=1}^{\infty}(2i+1)t_{2i+1}\partial_{t_{2i+1}}L,
\end{eqnarray}
which shows
\begin{equation}
\dfrac{\partial u_i}{\partial t^*_{1,1}}=2\left( x\dfrac{\partial u_i}{\partial x}+(i+1)u_i \right)+
2\sum\limits_{j=1}^{\infty}(2j+1)t_{2j+1}\dfrac{\partial u_i}{\partial t_{2j+1}},
\end{equation}
on account of
\begin{equation}
[x\partial,L]=-\partial+\sum\limits_{i=1}^{\infty}\left(x \dfrac{\partial u_i}{\partial x}+iu_i
  \right)\partial^{-i}.
\end{equation}
This is the end of the proof.\hfill $\square$
\begin{corollary}\label{evenadditionalflows}
From eq (\ref{additionalsymmeterygenerator}) we get $A_{1,l}=-lL^{l-1}$ when $l$ is even, thus
the corresponding flows on L are
\begin{equation}
\partial_{t^*_{1,l}}L=l[(L^{l-1})_{-},L]=\begin{cases}
0, \;\;for \; l=-2,-4,-6,\cdots.\\
\\
-l(\partial_{t_{l-1}}L), for \; l=2,4,6, \cdots.
\end{cases}
\end{equation}
\end{corollary}

  Motivated by the results on the KP hierarchy\cite{asv3, dl3} and the BKP hierarchy
  \cite{vdl2,tu1},
we can also define one generating function of additional symmetries
\begin{eqnarray}\label{generatingfunction}
Y^C(\lambda, \mu)&=&\sum\limits_{m=0}^{\infty}\dfrac{(\mu-\lambda)^m}{m!}
\sum\limits_{l=-\infty}^{\infty}\lambda^{-l-m-1}(A_{m,m+l})_{-}\notag\\
 &=&\sum\limits_{m=0}^{\infty}\dfrac{(\mu-\lambda)^m}{m!}
\sum\limits_{l=-\infty}^{\infty}\lambda^{-l-m-1}\left(
(M^mL^{l+m})-(-1)^{l+m}(L^{l+m}M^m)
\right)_{-},
\end{eqnarray}
which can be expressed by a simple form in the sequent proposition.
To this end, we need three well known and useful technique lemmas.
\begin{lemma}(\cite{dl1} \S 6.2.5)
For two pseudo-differential operators $P$ and $Q$,
the identity
\begin{equation}\label{PQidentity}
res_z[(Pe^{xz})(Qe^{-xz})]=res_{_{\partial}}[PQ^*]
\end{equation}
is true.
\end{lemma}
\begin{lemma}
Let $P$ be a pseudo-differential operators $P=\sum p_i\partial^i$, then
\begin{equation}\label{Pnegativepart}
P=\sum \partial^{i}\ \tilde{p}_i,\; \text{and}\
P_{-}=\sum\limits_{i=1}^{\infty}\partial^{-i}
res_{_{\partial}}(\partial^{i-1}P).
\end{equation}
\end{lemma}
\begin{lemma}(\cite{dl1} \S 6.3.2(ii)) If $f(z)=\sum\limits_{-\infty}^\infty a_iz^{-i}$, then
\begin{equation}\label{deltafunction}
{\mathrm res}_z[\zeta^{-1}(1-z/\zeta)^{-1}+z^{-1}(1-\zeta/z)^{-1}]
f(z)=f(\zeta).
\end{equation}
 (Here $(1-z/\zeta)^{-1}$ is understood as a
series in $\zeta^{-1}$ while $(1-\zeta/z)^{-1}$ is a series in
$z^{-1}$.)
\end{lemma}
\begin{proposition}\label{generatingfunctionadditionsymmetries}
\begin{equation}
Y^C(\lambda, \mu)=w(t,\mu)\partial^{-1}w(t,-\lambda)
+w(t,-\lambda)\partial^{-1}w(t,\mu)
\end{equation}
\end{proposition}
Proof. Starting from eq.(\ref{satorepresenation}) and
eq.(\ref{definitionadditionalflowsonphi}), and using the first two
lemmas at the first and second step, we get
\begin{eqnarray*}
\left(M^mL^{l+m} \right)_{-}&=&\sum\limits_{i=1}^{\infty}\partial^{-i}\ res_{_{\partial}}
[\partial^{i-1}\ \phi\ \Gamma^m\ \partial^{l+m}\ \phi^{-1}]\\
&=&\sum\limits_{i=1}^{\infty}\partial^{-i}\ res_{_{z}}
[
\left(\partial^{i-1}\ \phi\ \Gamma^m\ \partial^{l+m} e^{\xi(t,z)}\right)
\left((\phi^{-1})^*e^{-\xi{(t,z)}}\right)]\\
&=&\sum\limits_{i=1}^{\infty}\partial^{-i}\ res_{_{z}}
[
\left(\ z^{l+m}\partial^{i-1}\ \phi\ \Gamma^m e^{\xi(t,z)}\right)
\left(w^*(t,z)\right)]\\
&=&\sum\limits_{i=1}^{\infty}\partial^{-i}\ res_{_{z}}
[
\left(\ z^{l+m}\partial^{i-1}\ \phi\ (\partial_z^m e^{\xi(t,z)})\right)
\left(w^*(t,z)\right)]\\
&=&\sum\limits_{i=1}^{\infty}\partial^{-i}\ res_{_{z}}
[
\left(\ z^{l+m}\partial^{i-1}\  (\partial_z^m \phi e^{\xi(t,z)})\right)
\left(w^*(t,z)\right)]\\
&=&\sum\limits_{i=1}^{\infty}\partial^{-i}\ res_{_{z}} [ \left(\
z^{l+m}(\partial_z^m w(t,z))^{(i-1)} \right)
\left(w^*(t,z)\right)]\\
&=&res_{_{z}}[
\ z^{l+m}(\partial_z^m w(t,z))\ \partial^{-1}\
w^*(t,z)].
\end{eqnarray*}
Above,the final step is due to the identity $f\partial^{-1}= \partial^{-1}\ f-
\partial^{-1}\ f_x \ \partial^{-1}=\sum\limits_{i=1}^{\infty}\partial^{-i}\
f^{(i-1)}$ inversely. Here  $f_x=(\partial_x f)$ and $f^{(i-1)}=\dfrac{\partial^{i-1} f}
{\partial x^{i-1}}$, $f$ is a $C^{\infty}$ function in $x$.
Similarly,
\begin{equation*}
\left(L^{l+m}M^m\right)_{-}=res_z[(\partial_z^m\ z^{l+m}w(t,z))\ \partial^{-1}\ w^*(t,z)].
\end{equation*}
Taking them back into $Y^C$,which becomes
\begin{eqnarray*}
Y^C(\lambda,\mu) &=&\sum\limits_{m=0}^{\infty}\dfrac{(\mu-\lambda)^m}{m!}
\sum\limits_{l=-\infty}^{\infty}
\lambda^{-l-m-1}res_{_{z}}[
\ z^{l+m}(\partial_z^m w(t,z))\ \partial^{-1}\
w^*(t,z)]\\
&+&
\sum\limits_{m=0}^{\infty}\dfrac{(\mu-\lambda)^m}{m!}
\sum\limits_{l=-\infty}^{\infty}
(-\lambda)^{-l-m-1}
res_{_{z}}[(\partial_z^m\ z^{l+m}w(t,z))\ \partial^{-1}\ w^*(t,z)]\\
&=&res_{_{z}}[\sum\limits_{n=-\infty}^{+\infty} \dfrac{z^n}{\lambda^{n+1}}w(t,z+\mu-\lambda)
\ \partial^{-1}\ w^*(t,z)]\\
&+& res_{_{z}}[\sum\limits_{n=-\infty}^{+\infty}
\dfrac{(z+\mu-\lambda)^n}{(-\lambda)^{n+1}} w(t,z+\mu-\lambda)
\ \partial^{-1}\ w^*(t,z)]\\
&=&
w(t,\mu)\partial^{-1}w(t,-\lambda)
+w(t,-\lambda)\partial^{-1}w(t,\mu).
\end{eqnarray*}
Here the final step is due to eq.(\ref{deltafunction}). \hfill $\square$

Moreover,  proposition \ref{generatingfunctionadditionsymmetries} can also provide an explanation (from the view of
symmetry) for the form of the Lax operator of constrained the CKP hierarchy\cite{il1}.
To this end, let
\begin{equation}\label{eigenfunctionfromwavefunction}
\phi_1(t)=\int d\mu \rho_1(\mu)w(t,\mu), \;
\phi_2(t)=\int d\lambda \rho_2(\lambda)w(t,-\lambda),
\end{equation}
and $\rho_i$ are some suitable weighting function, then
they satisfy
\begin{equation}\label{eigenfunctionequation}
(\partial_{t_{2n+1}}\phi_i)=(B_{2n+1}\phi_i).
\end{equation}
Therefore, there exists symmetry reduction of the CKP hierarchy,
\begin{equation}
L^k=L^k_+ + \phi_1  \partial^{-1} \phi_2+ \phi_2  \partial^{-1} \phi_1,
\end{equation}
which is  the form of the Lax operator of the constrained CKP hierarchy\cite{il1}.
\section{String Equation}
    In corollary \ref{evenadditionalflows}, we did not mention the case when $l$ is odd.
This case is more interesting  because it is related to so called "string equation"\cite{douglas}
and thus deserves to be discussed separately with more details. In this section,
from now on, we set $l=2k$ and $k=1,2,3,\cdots$ then
\begin{eqnarray}
&A_{1,-(l-1)}=ML^{-(l-1)}+L^{-(l-1))}M=2ML^{-(l-1)}-(l-1)L^{-l},\label{generator1-of-addsymmwithstringeq}\\
&[A_{1,-(l-1)}, L^l]=-2l.\label{generator2-of-addsymmwithstringeq}
\end{eqnarray}
Therefore, we get a special action of the additional flows on $L^{l}$
\begin{eqnarray}\label{addflowswithstringeq}
\partial_{t^*_{1,-(l-1)}}L^{l}&=&[-(A_{1,-(l-1)})_{-},L^l]=[(A_{1,-(l-1)})_{+},L^l]+
[-(A_{1,-(l-1)}),L^l] \notag\\
&=&2[(ML^{-(l-1)})_+, L^l]+2l.
\end{eqnarray}
We can get the following proposition on the string equation.
\begin{proposition}\label{propstringequation}
If $L^l$ is a differential operator and it is independent of the additional variables
$t^*_{1,-(l-1)}$, then
\begin{equation}\label{CKPstringequation}
[L^l, \dfrac{1}{l}(ML^{-(l-1)})_+]=[L^{2k},\frac{1}{2k}ML^{-(2k-1)}-\frac{1}{2}
\frac{2k-1}{2k}L^{-2k}]=\textbf{1}
\end{equation}
is a string equation $[P,Q]=\textbf{1}$(P and Q  are two differential
operators) of the CKP hierarchy.
\end{proposition}
\noindent\textbf{Proof}
Setting $l$ as an even positive number into eq.(\ref{addflowswithstringeq}), and further
assuming that $L$ does not depend $t^*_{1,-(l-1)}$, we have $\partial_{t^*_{1,-(l-1)}}L^{l}=0$,which implies
$
\left[L^l, \dfrac{1}{l} (ML^{1-l} )_{+}\right]=1.
$
Moreover, $(A_{1,-(2k-1)})_{-}=0$ implies
$(ML^{-(2k-1)})_{-}=\frac{2k-1}{2}L^{-2k}$, and then $(ML^{-(2k-1)})_{+}=
ML^{-(2k-1)}-\frac{2k-1}{2}L^{-2k}$, which completes the proof. \hfill $\square$
\begin{corollary}\label{CKPcorstringeqautionthroughCKPflows}
The string equations  can be also expressed by
CKP flows as
\begin{equation}
-\frac{1}{2k}\sum\limits_{n\geq k+1}(2n-1)t_{2n-1}\Big( \partial_{t_{2n-(2k+1)}} L^{2k}
\Big)=1
\end{equation}
\end{corollary}
\noindent \textbf{Proof} By a direct calculation, the left hand side of eq.(\ref{CKPstringequation})
becomes
\begin{eqnarray*}
1&=&\left[L^{2k},\frac{1}{2k}\left(ML^{-(2k-1)}\right)_+\right]=
\left[L^{2k},\frac{1}{2k}\left(
\phi \sum\limits_{n=1}^{\infty}(2n-1)t_{2n-1}\partial^{2n-2k-1}\phi^{-1}
\right)_+\right]\\
&=&
\left[L^{2k},\frac{1}{2k}\left(
\phi \sum\limits_{n\geq k+1}^{\infty}(2n-1)t_{2n-1}\partial^{2n-2k-1}\phi^{-1}
\right)_+\right].
\end{eqnarray*}
Note that the change in the summation index is due to the identity
 $(\phi\partial^{-m}\phi^{-1})_+=0$,here $m$ is a positive integer.
 We also should note $L^{k}_+=(\phi\partial^k\phi)_+$ with $k\geq 0$, and then get
\begin{eqnarray*}
1&=&
\left[L^{2k},\frac{1}{2k} \sum\limits_{n\geq k+1}(2n-1)t_{2n-1}\left(L^{2n-2k-1}
\right)_+\right]=-\frac{1}{2k} \sum\limits_{n\geq k+1}
(2n-1)t_{2n-1}\left(\partial_{t_{2n-2k-1}}L^{2k}
\right).
\end{eqnarray*}
This completes the proof. \hfill $\square $

\begin{corollary}\label{CKPcorMLonelminuonenegativepart}
 If  $L^l$ and $M$ satisfy the string equations eq.(\ref{CKPstringequation}),
then
\begin{equation}
(ML^{-(l-1)})_{-}=\frac{l-1}{2}L^{-l}, (L^{-(l-1)}M)_{-}=-
\frac{l-1}{2}L^{-l}.
\end{equation}
Here $l=2k$ as before.
\end{corollary}

\noindent The $Q=\dfrac{1}{l}(ML^{-(l-1)})_+$ is of  infinite order.
Let assume that $t_i(i\ge q+l)$ are frozen, i.e. there exist only nonzero finite
independent variables $t_i(i\leq q+l)$,then $Q$ has order $q$.

 The string equations in eq.(\ref{CKPstringequation}) impose some restrictions on the Lax
operator of the CKP hierarchy. In general, we can proceed further to get more higher order
constraints on $L$  similar to what  Adler and Moerbeke \cite{av1} have done for the
KP hierarchy. To this purpose, we start
with a special generator,
\begin{equation}\label{CKPgeneratorAjplplusj}
A_{j,pl+j}=M^jL^{pl+j}+ L^{pl+j}M^j
\end{equation}
getting by the definition of generator of the additional symmetry
eq.(\ref{additionalsymmeterygenerator}), with $j=1,3,5,\cdots$, $p=-1,0, 1,2,3,
\cdots$.
The fact $pl+j$ is an odd number is used to get the form
of the  $A_{j,pl+j}$.
\begin{lemma}\label{CKPlemMjLplplusjnegativepart}
With same the conditions as proposition \ref{CKPpropVirasoroplushighervonLl},then
\begin{equation}
\left(M^jL^{pl+j} \right)_{-}=\begin{cases}
\prod\limits_{r=0}^{j-1} (\frac{l-1}{2}-r)\cdot L^{-l},\;\; p=-1,  \\
0, p=0,1,2,3,\cdots.
\end{cases}
\end{equation}
\end{lemma}
\noindent \textbf{Proof}
The corollary \ref{CKPcorMLonelminuonenegativepart} shows that the lemma holds for
$j=1$ and $p=-1$. Firstly, according to the induction method, we assume that the lemma is
verified for  a given  $j$ and $p=-1$,i.e.,
$\left((M^jL^{-l+j} \right)_{-}=\prod\limits_{r=0}^{j-1} (\frac{l-1}{2}-r)\cdot L^{-l}
$, and then  prove that the lemma is also true for this $j$, but $p\geq 0$. So let $p\geq 0 $, note that $L^{(p+1)l}$ is a differential operator, then
$$
\left( M^jL^{pl+j}\right)_{-} = \left( (M^jL^{j-l})_{-} L^{(p+1)l} \right)_{-}.
$$
Taking the assumption into it,
$$
\left( M^jL^{pl+j}\right)_{-} = \left( \prod\limits_{r=0}^{j-1} (\frac{l-1}{2}-r)\cdot L^{-l}
L^{(p+1)l} \right)_{-}=0
$$
by using the fact that  $L^{pl}$ is a differential operator. Secondly,
assume again that the lemma is true a given  $j$ and $p=-1$, and then  prove that the
lemma holds for the  $j+1$ and $p=-1$. To this end, by using identity $[M,L^j]=-jL^{j-1}$,
we compute
$$
\left(M^{j+1}L^{-l+j+1} \right)_{-}=\left(M^{j}ML^{j} L^{-l+1} \right)_{-}=
\left(M^{j}L^{j}M L^{-l+1}\right)_{-}-j\left( M^jL^{j-l} \right)_{-}.
$$
By using formula $M^jL^j=(M^jL^{j+p\cdot l}|_{p=0})_{-}=0$ we have just proven previously  and
the assumption, then
\begin{eqnarray*}
\left(M^{j+1}L^{-l+j+1} \right)_{-}&=&\left(M^{j}L^{j}(M L^{-l+1})_{-}\right)_{-}-
j\prod\limits_{r=0}^{j-1}(\frac{l-1}{2} -r)\cdot L^{-l}\\
&=&\frac{l-1}{2}\left(M^jL^{j-l} \right)_{-}-j\prod\limits_{r=0}^{j-1}(\frac{l-1}{2} -r)\cdot L^{-l}\\
&=&\prod\limits_{r=0}^{j}(\frac{l-1}{2} -r)\cdot L^{-l},
\end{eqnarray*}
as we expected. This completes the proof of the lemma.\hfill $\square$
\begin{lemma}\label{CKPlemLplplusjMjnegativepart}
With same the conditions as proposition \ref{CKPpropVirasoroplushighervonLl}, then
\begin{equation}
\left(L^{pl+j}M^j \right)_{-}=\begin{cases}
\prod\limits_{r=0}^{j-1} (-\frac{l-1}{2}-r)\cdot L^{-l},\;\; p=-1,  \\
0, p=0,1,2,3,\cdots.
\end{cases}
\end{equation}
\end{lemma}
\noindent \textbf{Proof}
The proof of this lemma is similar to the proof above. \hfill $\square$
\begin{proposition}\label{CKPpropVirasoroplushighervonLl}
Let $l=2k$ as before, $j=1,3,5,\cdots$, $p=-1,0, 1,2,3, \cdots$. If  $L^l$ satisfies the string
equation eq.(\ref{CKPstringequation})
and $L^l$ is a differential operator, then
\begin{equation}\label{CKPeqVirasoroplushighervonLl}
\partial_{t^*_{j,pl+j}}L^l=0
\end{equation}
\end{proposition}
\noindent \textbf{Proof}
According to the action  of additional symmetry flows on the Lax operator, i. e.
eq.(\ref{CKPadditionalflowsonLandM}),
and the explicit form of $A_{j,pl+j}$ in eq.(\ref{CKPgeneratorAjplplusj}),
\begin{equation}\label{eqVirasoroplushigherv}
\partial_{t^*_{j,pl+j}}L^l=-\left[\left(
M^jL^{pl+j}+L^{pl+j}M^j
\right)_{-}, L^l  \right].
\end{equation}
Taking the above two lemmas into it, the proposition is proved. \hfill $\square$

The equations in (\ref{CKPeqVirasoroplushighervonLl})are called Virasoro (for j=1) and higher
Virasoro constraints. As Dickey\cite{dl4} has done for the KP hierarchy, one can get the
explicit form of the Virasoro generators and its action on the  $\tau$ function from
eq.(\ref{CKPeqVirasoroplushighervonLl}). However, for the CKP hierarchy  this procedure
can not be performed due to the non-existence of a sole $\tau$-function.

\section{Conclusions and Discussions}
To summarize, we have constructed the additional symmetries in eq.(\ref{definitionadditionalflowsonphi})
together with proposition \ref{additionalsymmetrygenerator} of the CKP hierarchy,
and further showed that they form a new infinite algebra $W^C_{\infty}$ in proposition \ref{WinfiniteCalgebra}.
We have presented a simple expression  in proposition \ref{generatingfunctionadditionsymmetries}
for the generating function of the symmetries.  We have also derived the string equation
in proposition \ref{propstringequation} and the associated higher order constraints on $L$ in proposition \ref{CKPpropVirasoroplushighervonLl}.
Particularly, the string equations were  also given by the actions of  the CKP flows on $L^{l}$
in corollary \ref{CKPcorstringeqautionthroughCKPflows}. Our results show that CKP hierarchy has,
indeed, different properties related to additional symmetry comparing with the BKP hierarchy.

{\bf Acknowledgments}
{\small
This work is supported by the NSF of China under Grant No. 10301030 and
No. 10671187, and SRFDP of China. Support of the joint post-doc fellowship of TWAS(Italy) and CNPq(Brazil) at UFRGS is
gratefully acknowledged. J. H. thanks Professors Li Yishen, Cheng Yi(USTC,China) and
F. Calogero(University of Rome "La Sapienza", Italy) for long-term encouragements and supports.}



\end{document}